\documentclass[twocolumn,showpacs,preprintnumbers,amsmath,amssymb,prb,longbibliography]{revtex4-1}
\usepackage{soul}
\usepackage[bbgreekl]{mathbbol}
\usepackage{graphicx}
\usepackage{dcolumn}
\usepackage{bm}
\usepackage{mathtools}
\usepackage{hyperref}
\usepackage{color}
\graphicspath{{fig/}{img/}}

\DeclareMathAlphabet\mathbfcal{OMS}{cmsy}{b}{n}

\newcommand{\kn}{{\bf  k}}
\newcommand{\qn}{{\bf  q}}

\newcommand{\Gn}{{\bf G}}

\newcommand{\rn}{{\bf r}}

\newcommand{\zn}{{\bf z}}

\newcommand{\un}{{\bf u}}
\newcommand{\Jn}{{\bf J}}
\newcommand{\En}{{\bf E}}
\newcommand{\An}{{\bf A}}

\def\gsim{\lower.35em\hbox{$\stackrel{\textstyle>}{\textstyle\sim}$}}
\def\lsim{\lower.35em\hbox{$\stackrel{\textstyle<}{\textstyle\sim}$}}

\begin{document}
\title{Quantum Band Structure and Topology in  One Dimensional Modulated  Plasmonic Crystal}
\author{Luis Brey}
\affiliation{Instituto de Ciencia de Materiales de Madrid (CSIC), Cantoblanco, 28049 Madrid, Spain}
\author{H.A. Fertig}
\affiliation{  Department of Physics, Indiana University, Bloomington, IN 47405 and \\
Quantum Science and Engineering Center, Indiana University, Bloomington, IN, 47408}
\begin{abstract}
Band structures of electrons in a periodic potential are well-known to host topologies that impact their behaviors at edges and interfaces.  The concept however is more general than the single-electron setting.  In this work, we consider topology of plasmons in a two-dimensional metal, subject to a unidirectional periodicity.  We show how the plasmon modes and wavefunctions may be computed for such a periodic system, by focusing on the confined, quantized photon degrees of freedom associated with the plasmon modes.  At low frequencies the plasmons disperse with wavevector as $\sqrt{q}$; however at higher frequencies one finds a series of bands and gaps in the spectrum.  For a unidirectional periodic electron density with inversion symmetry, we show that each band hosts a Zak phase $\gamma_n$ which may only take the values $0$ or $\pi$.  Each gap has a topological index $\nu$ that is determined by the sum of the Zak phases below it.  When the system has an interface with the vacuum, one finds in-gap modes for gaps with non-trivial topologies, which are confined to the interface.  In addition, interfaces between systems that are the same except for their topologies -- which can be created by a defect in the lattice in which half a unit cell has been removed -- host in-gap, confined states when the topological indices of the relevant gaps are different.  We demonstrate these properties numerically by analyzing a Kronig-Penney-type model in graphene, in which the electron density is piecewise constant, modulating between two different densities.  In addition, we consider a related plasmon system modeled after the Su-Schrieffer-Heeger (SSH) model.  We show that the topological phase diagram of the lowest energy bands is highly analogous to that of the SSH tight-binding system.
\end{abstract}
\date{\today}

\maketitle

\section{Introduction}
Plasmons are collective excitations of an electron gas that appear due to the long-range character of the Coulomb interaction between electrons \cite{Pines:1952aa,Pines:1956aa,Sawada:1957aa,Vignale-book,Pitarke:2007aa}.
When the electrons are confined to move in two-dimensions the plasmons become gapless and disperse as $\sim \sqrt{q}$, where ${\bf q}$ is the two-dimensional (2D) momentum \cite{Ritchie:1957aa,Ando:1982aa,Stauber:2014aa,Stauber:2013ac,Stauber17}.  Plasmons in general may be described in different ways.  For example, one may focus on the charge degrees of freedom of a conducting system,
so that they arise as divergences in its charge density response function, thus corresponding to self-sustained charge density excitations \cite{Vignale-book}. In 2D electron gases, they can alternatively be understood as electromagnetic waves confined to the vicinity of the electron sheet.  In this context, they are also known also  as surface-plasmon polaritons \cite{nikitin-book,Nikitin:2011aa,Koppens:2011ab,Slipchenko:2013aa,Peres-book}. The latter approach is particularly useful for describing coupling of plasmons with atoms \cite{Gonzalez-Tudela:2011aa,Torma:2015aa}, resulting in plasmon-mediated interactions among quantum emitters (as realized by the atomic electronic degrees of freedom) \cite{Gonzalez-Tudela:2013aa}. The plasmon-polariton approach is also useful for understanding chemical reactions driven by confined electromagnetic fields \cite{Fregoni:2022aa}.

Plasmons confined in graphene are particularly interesting.  Their frequencies, which fall within the mid-infrared and terahertz spectral ranges, can be tuned via gating \cite{Jablan:2009aa,Grigorenko:2012aa,Fei:2012ab}. These plasmons can be directly observed in near-field microscopy experiments \cite{Chen:2012aa,Fei:2012aa,Fei:2013aa} and support long propagation lengths \cite{Woessner:2015aa}.
Moreover, graphene can be patterned or gated to modulate the density of carriers and create periodic nanostructures in one \cite{Nikitin:2012aa,Zhao:2015aa,Dias:2017aa,Rappoport:2021aa,Guo:2023aa,nikonov2024plasmonstwodimensionalelectronsystems,Ju:2011ab,Strait:2013aa,Bylinkin:2019aa} and two \cite{Jin:2017aa,Azar:2020aa,Xiong:2019aa} dimensions.
These metamaterials, in analogy to photonic crystals \cite{Joannopoulos:08:Book}, are referred to as plasmonic crystals. The periodicity induces destructive interference, potentially creating one or more frequency gaps where light cannot propagate.

In analogy with electronic systems \cite{Vanderbilt:2018aa}, the plasmon band structure of a plasmonic crystal does not contain all the information about the modes.  In particular,
the  topology of the plasmon wave functions can be crucial for understanding the localization behavior of plasmons at defects and interfaces.
Topological  and geometric effects of the type now well-known for electron bands \cite{Vanderbilt:2018aa} have been proposed to occur in plasmonic \cite{Guo:2017aa,Cuerda:2024aa,Cuerda:2024ab} and photonic crystals \cite{Lan:2022aa,Wu:2015aa,Mingaleev:2003aa,Gupta:2022aa}.

In graphene, one-dimensional plasmonic crystals can be created by periodically varying a metal-gated structure, or by depositing a periodic metal grating on the graphene sheet. This modulation induces a periodic variation in the carrier density profile, perturbing the plasmons and generating subbands \cite{Ju:2011ab,Strait:2013aa,Brey:2020ac}.
The simplest description of this structure is a Kronig-Penney \cite{Kronig:1997aa} (KP) model, consisting of alternating regions with different charge densities.
This one-dimensional approach captures the essence of band structure formation in such metamaterials.
Most calculations of plasmon band structures in such metamaterials have been performed classically, matching the electric and magnetic fields  at the interfaces between regions with different densities using appropriate boundary conditions\cite{Nikitin:2011aa,Peres-book}.

For non-interacting electrons, a paradigm for topological band structures is the Su-Schrieffer-Heeger (SSH) \cite{Su:1979aa} model.  In its simplest realization, this is a tight-binding structure with two sites per unit cell and two different hopping amplitudes, resulting in two bands which may or may not be topological in nature.
Because of its simplicity, most existing work on the topology of one-dimensional crystal plasmons focus on describing the two lowest subbands, mapping the system onto a 2$\times$2 tight-binding SSH-like Hamiltonian \cite{Rappoport:2021ab,Liu:2022aa,Smith:2021aa,miranda2024topologyonedimensionalplasmoniccrystal}.
This can be achieved, for example, by building the unit cell to contain two equal regions of high electron density, separated by two unequal regions of low electron density \cite{Rappoport:2021ab,Liu:2022aa,Smith:2021aa}.
However, for this mapping to be valid, 
the energy gap between the second and third subbands should  be larger than the gap between the first and second subbands \cite{Pedersen:1991aa,Gupta:2022ab} if one wishes to ignore higher energy subbands.
In general, a continuous modulation of the density results in the formation of an infinite number of bands separated by similar-sized gaps.  Such situations cannot be treated analytically or fully analyzed topologically within the framework of a tight-binding Hamiltonian.

In this work, we present a method for computing  the plasmon band structure and wavefunctions  of a doped Drude-like graphene layer in the presence of a periodic continuous modulation of the charge density.
We apply this method to the case of a one-dimensional KP-like modulation of the Fermi energy.
Our focus is on unidirectional plasmonic waveguides, and we analyze in detail plasmons with zero momentum in the direction transverse to the modulations, where one expects to find the lowest energy modes.
By calculating the band structure and analyzing
the symmetry of the wavefunctions, we determine the Zak phases of the plasmon subbands and the \textit{Z}$_2$ topological invariants of the system.

The remainder of this paper is organized as follows.  In Section II we present a quantum Hamiltonian that describes plasmons in terms of confined electromagnetic modes, as well as by their
periodic modulation of the Fermi energy. In Section III, we explore the plasmon band structure when the graphene  Fermi energy exhibits a one-dimensional KP-like modulation. Section IV discusses the Zak phases of the plasmon subbands and their relationship with the centers of the Wannier functions of these subbands. Additionally, we analyze how the non-trivial topology of the plasmon subbands leads to a topological \textit{Z}$_2$ invariant for the system when the energy lies within a subband gap.
At the interface between gapped systems with different \textit{Z}$_2$ invariants, interface states with energies in the middle of the gap are expected to emerge. In Section V, we demonstrate the existence of such states at the interface between two distinct topological plasmonic crystals, as well as at the interface between a plasmonic crystal and a plasmon vacuum.
In Section VI, we examine a KP-like model in which the unit cell consists of two equal regions of high electron density separated by two unequal regions of low density. We demonstrate that this system shows the same topological properties as the SSH model.
We conclude in Section VII with a summary of our results.












\section{Plasmonic Crystals.}
At  long wavelengths  and low energies, plasmons in two-dimensional metals  depend  only on the local optical conductivity $\sigma (\omega; E_F)$ of the electron gas.  For frequencies higher than the  continuum of the intraband electron-hole pair  energies, the metal is essentially dissipationless, and is  characterized by a local optical conductivity which for long wavelengths and small frequencies has the Drude-like form \cite{Wunsch:2006aa,Hwang:2007aa,Brey:2007aa} $\sigma (\omega; E_F)$=$ i \frac {D} {\omega}$, where $D=\frac {e ^2 E_F}{\hbar ^2 \pi}$ is the Drude weight and $E_F$ is the Fermi energy.
In graphene the band structure disperses linearly with momentum, and  the  Fermi energy is related to the  carrier density, $n_0$, by $E_F=\hbar v_D \sqrt{\pi n_0}$,   with $v_D$ being the speed of electrons at the graphene Dirac points \cite{Guinea_2009, Katsnelson-book, Stauber:2014aa}. In the semiclassical limit the plasmon frequency, $\omega _q $=$\sqrt{ \frac { D}{2 \epsilon _d \epsilon _0} q}$ (with $\epsilon _d $ the dielectric constant of the surrounding medium), is obtained from Maxwell's equations, with proper matching of the fields across the 2D metal sheet \cite{nikitin-book,Brey:2020ac}.

Such semiclassical plasmons can be quantized in the
near -field approximation, which is appropriate when the wavelength of the plasmons is much smaller than that of free light at the same frequency. In this approximation  only the longitudinal electric field is considered, and the magnetic field associated with the plasmon mode is neglected. The quantum Hamiltonian describing plasmons in a uniform sheet of a doped metal that emerges from this approach is \cite{ Elson:1971aa,Gruner:1996aa,Archambault:2010aa,Hanson:2015aa,Ferreira:2020aa,Brey:2024aa}
\begin{widetext}
\begin{equation}
\hat H   =     \frac {  \epsilon _0 \epsilon _d} 2      \int  \int d \rn dz   {\hat \En} (\rn,z) {\hat \En} (\rn,z)  +
\frac 1 2   \int  \int d \rn dz   D \delta (z)    {\hat \An} (\rn,z) {\hat \An} (\rn,z)=\sum_{\qn} \frac {\hbar \omega _q } 2 \left ( \hat a _{\qn} \hat a ^{\dagger}_{\qn} + \hat a ^{\dagger}_{\qn} \hat a _{\qn} \right ),
\label{QH}
\end{equation}
\end{widetext}
where the operator $\hat a _{\qn}^{\dagger} $ creates a plasmon with 2D momentum $\qn$ and frequency $\omega _{q}$.  In terms of $\hat a _{\qn}$, $\hat a _{\qn}^{\dag}$ the electric field and vector potential operators are
\begin{eqnarray}
{\hat \En} (\rn,z)  &
=& \sum _{\qn}    \sqrt{ \frac {\hbar \omega _q }{ 2 \epsilon _0 \epsilon_d S  }}   e ^{i \qn  \rn } \un (\qn,z)  \hat a _{\qn} + h.c., \nonumber \\
{\hat \An} (\rn,z) & = &  -i \sum _{\qn}\sqrt{ \frac {\hbar }{ 2 \epsilon _0 \epsilon_d S \omega _q  }}   e ^{i \qn  \rn } \un (\qn,z)  \hat a _{\qn} + h.c., \nonumber \\
\label{QH1}
\end{eqnarray}
where $S$ is the sample area and the vectors $\un (\qn,z)$ are given by
\begin{equation}
{\bf u}(\qn,z)=   e^{-q |z|} \sqrt { \frac q  2 } \left (i \frac {\qn } q -  \frac {z}{|z|}
 \hat {\zn} \right )  \, .
 \label{QH2}
\end{equation}
The first term in
Eq. \ref{QH} represents  the energy stored in the electric field  and the second term, which is nonzero only in the conducting layer, is the kinetic energy corresponding to the motion of the charge
carriers. The Drude weight represents the stiffness of the electron gas against oscillations.

A non-uniform 2D system is defined by a density of charge that modulates in space.  Plasmons in such systems are conventionally studied in a local approximation \cite{Peres:2012aa,Slipchenko:2013aa,silveiro:2013aa,Beckerleg:2016aa,Huidobro:2016aa,Huidobro:2016ab,nikitin-book,Brey:2020ac}, that considers the optical conductivity at each point in space to be determined by the local Fermi energy. This in turn is obtained from the
local charge density using the Thomas-Fermi approximation \cite{Brey:2009aa}, $\sigma (\rn)$=$\sigma (E_F[n(\rn)] )$, which is implemented by endowing the Drude weight with a spatial dependence. When this dependence is periodic in space, the kinetic energy part of the Hamiltonian
Eq.\ref{QH1} is modified by replacing the uniform Drude weight $D$ by
\begin{equation}
D(\rn) = D_0 +\delta D (\rn)=D_0 + \sum_  {\Gn \ne 0 } D_{\Gn  } e ^{i \Gn \rn} \, ,
\label{DR}
\end{equation}
where $D_0$ is the average Drude weight, $\{ \Gn \}$ are the reciprocal lattice vectors associated with the periodicity of the modulation, $D_{\Gn}$=$\frac 1{S_{u.c.}} \int_{u.c.} d^2 r D(\rn) e ^{-i \Gn \rn}$ and $\int_{u.c.} d^2 r$ and $S_{u.c.}$ are a unit cell integral and the unit cell area, respectively.
Using
this form of $D(\rn)$ and the definition of the vector potential operators as functions of
$\hat a _{\qn}^{\dagger} $  and  $\hat a _{\qn}$, Eq. \ref{QH1}, one arrives at
\begin{equation}
\hat H \! = \!  \sum _{\kn,\Gn} \hbar \omega _{|\kn+\Gn|}  \left ( \hat a^{\dagger}  _{\kn + \Gn} \hat a _{\kn+ \Gn }  \! + \!  \frac 1 2 \right ) \! + \!
\sum_{\kn,\Gn,\Gn'}
 \hat V (\kn \! + \!\Gn,\kn \! + \! \Gn '), \label{HTotal}
\end{equation}
with
\begin{widetext}
\begin{equation}
\hat V (\kn+\Gn,\kn+\Gn ')    = V (\kn+\Gn,\kn+\Gn ')
\left ( \hat a ^{\dagger} _{\kn + \Gn} \hat a _{ \kn +\Gn '} - \hat a  _{-\kn -\Gn ' } \hat a _{\kn+\Gn } + h.c. \right ) \, ,
\label{Pert}
\end{equation}
where
\begin{equation}
V (\kn+\Gn,\kn+\Gn ') =  \frac 1 4  \hbar \sqrt{  \omega _{|\kn+\Gn|}    \omega _{|\kn+\Gn '|}   } \,
\frac { (\kn+\Gn) \cdot (\kn+\Gn ') }  { |\kn+\Gn||\qn+\Gn '| }  \frac {D_{\Gn- \Gn '}} {D_0} \, .
\label{pote}
\end{equation}
\end{widetext}
In Eq. \ref{HTotal},  the momentum $\kn$  is restricted to values in the first Brillouin zone associated with the periodicity of the Drude weight.
The first term appearing in Eq. \ref{Pert} is a resonant contribution that describes the transfer of  momentum $\Gn- \Gn '$ between two plasmons. The second term describes processes which are non-conserving in the number of plasmons,  describing the simultaneous annihilation of a pair of plasmons while transferring  a momentum
$\Gn- \Gn '$ to the system. In quantum optics, such non-conserving contributions to the Hamiltonian are known as counterrotating (CR) terms \cite{Grynberg-book,Vogel-Welsch-Book,Kavokin-Book,Torma:2015aa,Frisk-Kockum:2019aa,Kirton:2019aa}.
Using this quantum formalism we obtain in a transparent way the
the scattering between two plasmons with momentum $\qn$ and $\qn'$ due to density inhomogeneities \cite{Sziklas:1965aa,Rudin:1993aa,Torre:2017aa,Cao:2021aa,Potential}.
The scattering potential of plasmons by a spatial modulation of the Drude weight is formally equivalent to the Rayleigh scattering of phonons by a modulation of the atomic mass \cite{Ziman_book}.


The  Hamiltonian Eq. \ref{HTotal} is bilinear in field operators and can be we diagonalized using a Bogoliubov-Hopfield  symplectic transformation \cite{Hopfield:1958aa,Ciuti:2005aa,Brey:2024aa}.  The transformed Hamiltonian takes the form
\begin{equation}
\hat H = \sum _{\kn,n}  E_{k,n} \left (\hat b ^{\dagger} _{\kn,n}  \hat b _{\kn,n} + \frac 1 2 \right )
\end{equation}
where
$E_{\kn,n}$ are the energies of the plasmon modes, where here  $n$ is a band index.
The eigenstates corresponding to a mode $\kn,n$ have the form
\begin{equation}
\hat b ^{\dagger}  _{n,\kn} =\sum _{\Gn} \left( \alpha ^n _{\kn+\Gn} \hat a ^{\dagger} _{\kn+\Gn} +\beta ^n_{\kn+\Gn} \hat a  _{-\kn -\Gn} \right).
\end{equation}
The counterrotating terms in Eq. \ref{Pert} lead to an admixture of creation and annihilation operators of the uniform system forming the operators that create and annihilate normal modes of the periodic system.
The operators $\hat b$ and $\hat b ^{\dagger}$ are bosonic, and order to satisfy the proper commutation relations, the coefficients $\{ \alpha \}$ and $ \{ \beta \}$ must satisfy the conditions
\begin{equation}
\sum_{\Gn}
\left[ ({\alpha ^n   _{\kn+\Gn} } )^2 -(\beta ^{n }_{\kn+\Gn})^2 \right] =1.
\label{norma}
\end{equation}

Real space wave functions for the crystal plasmons can be cast in a ``particle-hole'' spinor form,
 \begin{widetext}
 \begin{equation}
{ \Psi}_{n,\kn} ({\bf r})=
 \left  (
 \begin{array} {c}
 \psi ^{p}
 _{n,\kn} (\rn)\\ \psi ^{h} _{n,\kn} (\rn) \end{array} \right )
 =
 \left (
 \begin{array} {c}
 \langle \Phi _0 | \hat \Psi (\rn) \hat b  ^{\dagger} _{n,\kn} |\Phi _0 \rangle
 \\
 \langle \Phi _0 | \hat \Psi ^{\dagger} (\rn) \hat b  ^{\dagger} _{n,\kn} |\Phi _0 \rangle
 \end{array} \right ) =
  \frac 1 {\sqrt{S}}
 \left (
 \begin{array} {c}
 \sum _{\Gn} \alpha ^n _{\kn+\Gn} e ^{i(\kn+\Gn) \cdot \rn} \\
  -\sum _{\Gn} \beta ^n _{-\kn-\Gn} e ^{i(\kn+\Gn) \cdot \rn}
 \end{array} \right )\, .
 \label{realwf}
 \end{equation}
 \end{widetext}
Here $\hat \Psi (\rn)$=$\frac 1 {\sqrt{S}} \sum _{\qn} e ^{i \qn r }\hat a _{\qn}$ is an operator than annihilates a plasmon at location $\rn$, and
$|\Phi_0\rangle$ is the vacuum of the plasmonic crystal, defined by $b_{n,\kn} |\Phi_0 \rangle$=0, for all $n$ and $\kn$.

The modulation of the Drude weight couples
plasmons with momenta $\qn$ and $\qn +\Gn$.   At the center  and edges  of the Brillouin zone, unperturbed plasmons with momenta $\qn$ and $\qn +\Gn$ become degenerate and thus are strongly coupled by the perturbation \cite{Harrison_book,Aschroft_book,Cardona-book}. This  coupling  opens gaps in frequency at which plasmons cannot propagate.

\section{One-dimensional  modulation.}

\begin{figure}
\includegraphics[width=8cm]{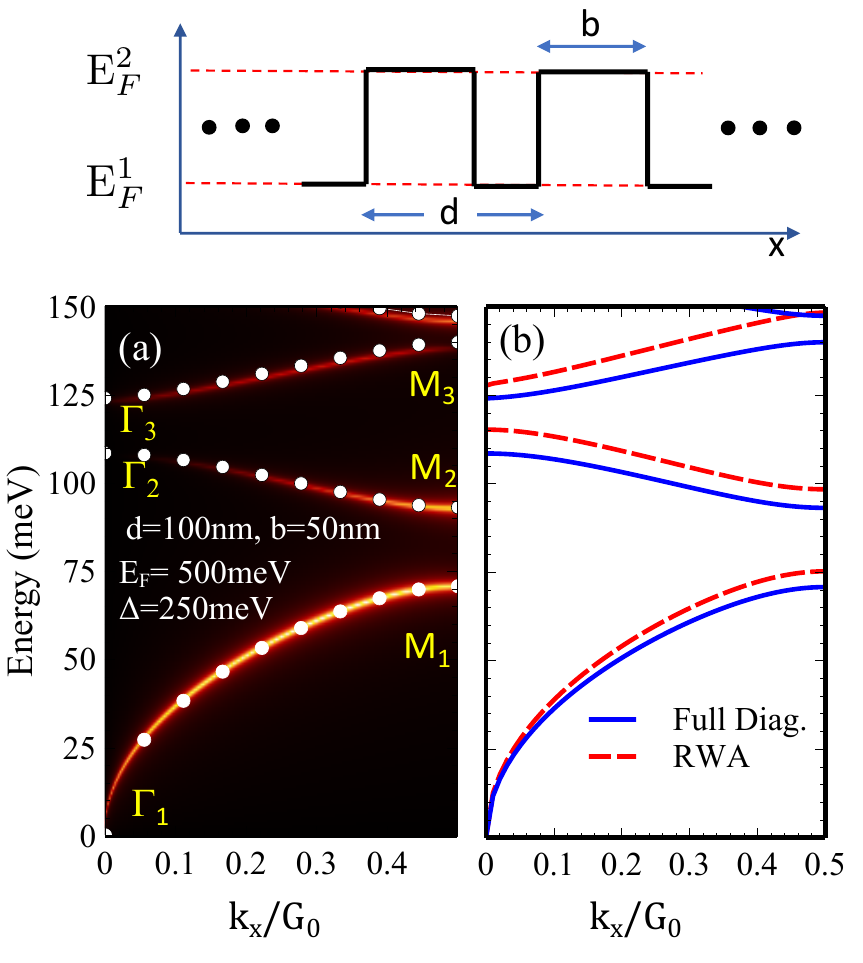}
\caption{
Upper panel: Schematically diagram of one-dimensional
modulation of  the Fermi energy  in a Kronig-Penney  model.
Bottom panels:
Plasmon band structure for a 1D modulation of period  $d$=100nm. The unit cell consists
of two equal regions with Fermi energies
$E_F ^{(1)}$=250meV and $E_F^{(2)}$=750 meV, and dielectric constant $\epsilon _d$=5.  In (a), the  white points correspond to the dispersion obtained by diagonalizing Eq. \ref{HTotal}. The color plot indicates the loss function of  the same system as obtained semi-classically. In  (b) we compare the dispersion  obtained by diagonalizing the full Hamiltonian Eq. \ref{HTotal} with that obtained in the RWA, which neglects terms that are non-conserving in plasmon number.
The states denoted as $\Gamma_n$ and $M_n$ represent the $n^{th}$ state at the $\Gamma$ and $M$ points, respectively.
}
\label{Plasmon_band}
\end{figure}

In this section, we analyze the plasmon band structure of a system with a one-dimensional  modulation of the Drude weight of period $d$.
The reciprocal lattice vectors are $G_n=n G_0$, with $G_0=\frac {2 \pi}d$ and $n$ an integer, and we write the spatially varying Fermi energy
in the form $E_F(x)=\sum_{G_n}E_{F,n} e^{iG_n x}$.
As a simple model we analyze a Kronig-Penney model in which regions of width $b$ having a Fermi energy $E_F^{(2)}$= $E_F$+$\Delta E_F$ alternate with regions of width $d$-$b$ having Fermi energy $E_F ^{(1)} $= $E_F$-$\Delta E_F$.  (See Fig. \ref{Plasmon_band}.)

Following the strategy described in the previous section, we can obtain the plasmon band structure by diagonalizing the resulting Hamiltonian (Eq. \ref{HTotal}.)  The white dots in
Fig. \ref{Plasmon_band}(a) show the result for unit cell parameters
$d$=100nm, $b$=50nm, $E_F$=500meV and $\Delta E_F$=250meV. To verify the accuracy of this quantum calculation, we also plot the loss function of this system obtained through self-consistent solutions of the  Poisson equation, Ohm's law, and the continuity equation. (Details are provided in Appendix A).  The agreement is excellent, which is perhaps not surprising since the Hamiltonian Eq. \ref{HTotal} describes a system of coupled harmonic oscillators, for which classical and quantum modes have identical frequencies.
In Fig. \ref{Plasmon_band}(b) we
compare the dispersion of the full Hamiltonian Eq. \ref{HTotal} with what is obtained in the Rotating Wave Approximation (RWA), which consists of neglecting the counter-rotating terms in the Hamiltonian,
Eqs. \ref{HTotal}-\ref{Pert}.
The RWA shifts the bands to higher energies, with the shift becoming more significant as the modulation of the Drude weight increases. Throughout the remainder of this work, unless stated to the contrary, we will include the counter-rotating terms in our calculations.  Finally, we note that, because
the Fermi energy is non-vanishing across the entire unit cell, at low energies the modes disperse as $\sqrt q$, indicating that the periodic modulation has little effect at long wavelengths and low frequencies.

The modulation of the Drude weight  creates  energy gaps  both  at the Brillouin zone center [$\Gamma$ point, $\kn = (0,0)$] and  at its edge [$M$ point, $\kn = (\frac {G_0} 2, 0) $].
The gap that separates the two lowest energy subbands occurs at the $M$ point, where the states of the unperturbed system $|\kn=-\frac {G_0}2\hat{x} \rangle$ and $|\kn=\frac {G_0}2 \hat{x}\rangle$ are degenerate, and are strongly admixed by the modulation of the Drude weight (see Eq. \ref{pote}.)
To linear order in $V$, the lowest energy crystal plasmon modes at the $M$ point may be
obtained by evaluating Eq. \ref{HTotal} in the subspace of these two (unperturbed) states.
This results in the effective $2 \times 2$  effective Hamiltonian \cite{Aschroft_book}
\begin{equation}
H \! = \!
\left (  \begin{array} {cc} \hbar \omega _{\frac {G_0} 2} & 0   \\
0                                & \hbar \omega _{\frac {G_0} 2}
\end{array} \right ) \! + \!
\left (  \begin{array} {cc} 0 & - \frac   1 4  { \hbar \omega _{\frac {G_0} 2} }   \left(\frac {E_{F,1} }{ E_{F,0}} \right)  \\
- \frac   1 4  { \hbar \omega _{\frac {G_0} 2} }  \left( \frac {E_{F,1} }{ E_{F,0}} \right) & 0 \end{array} \right ) \, .
\label{Eff_H}
\end{equation}
The eigenvalues of Eq. \ref{Eff_H}  are
$$\hbar\omega_{\pm} = \hbar  \omega _{\frac {G_0} 2} (1  \! \pm \! \frac 1 4 \left | \frac {E_{F,1} }{E_{F,0}} \right | ),  $$ and its eigenvectors are
$$
|\Psi_{\pm,M} \rangle = \frac 1 {\sqrt {2}} \left ( |\kn=-\frac {G_0 }2 \hat{x}) \rangle  \mp {\rm sgn}  \left (E_{F,1} \right ) |\kn=\frac {G_0 }2 \hat{x}\rangle \right ).$$ Note that while the energy gap is independent of the sign of $E_{F,1}$,  the parity of the wave function depends upon it. This fact plays a role in the band topology, as we now discuss.

\section{Zak Phase in one-Dimensional Plasmonic Crystals.} \label{Zak1D}

The band structure of electrons in a crystal by itself does not contain all the information about the electronic structure and the accompanying electron dynamics in the system. Modern band theory shows that the {\it topology} the of electron states is also necessary for understanding and predicting various properties.  In particular, for one-dimensional systems, the presence of localized states at defects and sample boundaries depends on this topology \cite{RevModPhys.82.1959,Vanderbilt:2018aa}.

The topological content of energy bands can often be expressed in terms of Berry phases.
Quite generally, the Berry phase is a geometrical global phase that a quantum state vector, specified by some set of parameters, accumulates as it is adiabatically transported around  a closed path in the parameter space.
In band theory the relevant vector spaces consist of the Bloch wave functions, and the parameter space is formed by the wavevectors themselves \cite{Vanderbilt:2018aa}. However, the calculation of Berry phases involves  computing inner products of wave functions at different wavevectors, for which the relevant inner products of the Bloch wavefunctions $\langle\Psi_{n,\kn_1}|\Psi_{n,\kn_2}\rangle$ vanish for $\kn_1 \ne \kn_2$ but not for $\kn_1 = \kn_2$. This means they do not have the smoothness needed to reveal phase changes along a continuous path in momentum space.  However, this property {\it is} present for spinor-like cell-periodic  functions $U_{n,\kn} (\rn)$=$ e^{-i \kn \cdot \rn} \Psi_{n,\kn} (\rn)$. The improved behavior results from the fact that these functions
satisfy the same boundary conditions for all wavevectors $\kn$, so that they may be mapped to a single Hilbert space in which states evolve smoothly with $\kn$.

In one-dimensional systems, the Berry phase acquired by wavefunctions of an isolated band, $n$, as the wavevector $k$  adiabatically  evolves
through the entire Brillouin zone, is known as the Zak phase \cite{Zak:1982aa,Zak_1989}.  In electronic tight-binding systems,
the scalar inner  product of the wave functions that are needed to compute the Zak phase have the form $V _{n,k}^{\dag} \cdot V_{n,k '}$, where the entries of $V_{n,k }$ are associated with amplitudes for an electron to be in different orbitals/sites.  Here we are interested in finding a corresponding expression for the Zak phase for plasmons in a one-dimensional crystal structure.  Because the plasmon states correspond to coherent admixtures of {\it bosonic} operators that add and remove plasmons from the corresponding uniform system, the needed inner products
should be modified
\cite{Shindou:2013aa,Goren:2018aa} to  $\langle U _{n,k } | \sigma _z | U_{n, k'}\rangle$  (see Eq. \ref{norma}.)  Note that the scalar product in this expression involves an integral over a single unit cell.
The Zak phase  of an isolated band in a plasmonic crystal then takes the form
\begin{equation}
\gamma _{n} = i \int _{-\frac {\pi} d} ^{\frac {\pi} d} dk \langle U_{n,k} | \sigma _z |  \partial _k U _{n,k} \rangle.
\end{equation}

In further analogy with electron systems, it is possible define Wannier functions for isolated plasmonic bands through the relation
$w _{n,X}(x)= \frac d {2\pi} \int _{-\frac{\pi} d } ^{\frac {\pi } d } e^{-i k X} \Psi _{n,k} dk$, where the allowed values of $X$ are the unit cell centers of the crystal.
The location of the Wannier center in the $X=0$ unit cell of an isolated band is then given by
\begin{equation} \bar x _n =  \int _{-\frac d 2} ^{\frac  d 2 } dx \,
w ^\dag _{n,X=0}(x) \, \sigma _z \, x \, w  _{n,X=0}(x),
\end{equation}
and is proportional to the Zak phase of the isolated band $n$,
\begin{equation}
\bar x _n = \frac d {2 \pi} \gamma_n \, .
\label{xntogamma}
\end{equation}

In general,
both the Zak phase and the Wannier center can assume any value. However, if the system possesses spatial inversion symmetry,
constraints on the Wannier center emerge that are topological in nature.  For one-dimensional periodic potentials such as we consider, inversion-symmetric systems always have
two inequivalent inversion centers, and the Hamiltonian will be inversion-symmetric when one of these is at the origin ($x=0$), in which case the other is at the unit cell edge ($x$=$d/2$). In analogy with the corresponding electron system \cite{Zak:1982aa}, the plasmon Wannier functions must be centered at one of these points, so that $\bar x _n$ can only take the values $0$ or $d/2$.
Therefore, in a plasmon crystal whose Hamiltonian has spatial inversion symmetry, the Zak phase is a topological index which can only have the values $0$ or $\pi$. Importantly, \textit{which} of these values a given band takes depends on which of the inversion symmetric points is located at the origin.
Because of this,  each isolated band $n$ may have one of two different topological characters, $\gamma _n$=0 or $\gamma_n=\pi$.  In what follows we label these two topologies
$D_1$ and $D_2$, respectively.  For concreteness, in our system (the Kronig-Penney plasmonic crystal), we take $D_1$  to correspond to taking the origin at the center of the region with higher Fermi energy, whereas in $D_2$, the origin is taken at the center of the region with lower Fermi energy.

In systems with spatial inversion symmetry, Zak phases turn out to be related in a simple way to the parity of the wavefunctions at high symmetry points of the Brillouin zone. Specifically, one may show \cite{Fu:2007aa,Hughes:2011aa,Miert:2017aa}
\begin{equation}
e^{i \gamma _n }=   \xi _n (\Gamma) \, \xi _n (M),
 \label{gamma}
 \end{equation}
where $\xi _n (\kn)$ represents  the parity of the wavefunction $\Psi_{n,\kn}$ at wavevector $\kn$, and $\Gamma$ and $M$ are the time reversal invariant momenta for this system.  We demonstrate Eq. \ref{gamma} in Appendix B.
More globally, for
a gap between subbands $n$ and $n+1$, one may construct \cite{Fu:2006aa,Vanderbilt:2018aa} a $\textit{Z}_2$ topological invariant, $\nu$,  which depends on the Zak phases of the bands with band indices less than or equal to $n$ and is defined as
\begin{equation}
(-1) ^{\nu} = \prod _{ i \le n}
\xi _i (\Gamma) \, \xi _i (M)
\label{Chern_index}
\end{equation}
As we shall see, these topological invariants dictate the behavior of the spectrum when a system is cut at an inversion-symmetric point, and when there is an interface between systems of $D_1$ and $D_2$ topology.

 \begin{figure}
\includegraphics[width=8cm]{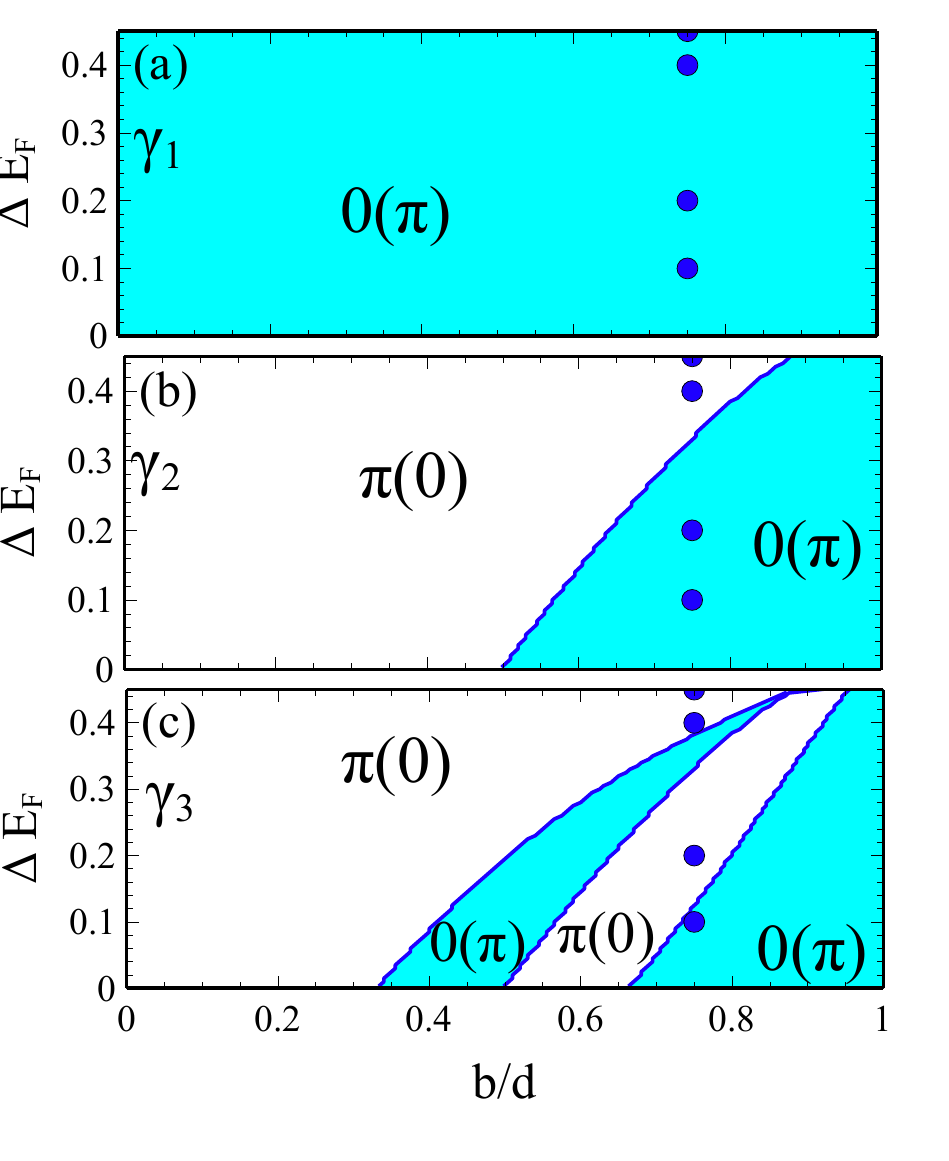}
\caption{
Zak phases of the three lowest energy subbands: (a) $\gamma _1$, (b) $\gamma _2$ and (c) $\gamma_3$, for a Kronig-Penney like model  (see Fig. \ref{Plasmon_band}) with $d$=100nm, $E_F$=500meV and different values of $b/d$ and $\Delta E_F$. $E_{F}^{(1,2)}$=$E_F\pm \Delta E_F$.  The phases correspond to topology $D_1$. The Zak phases corresponding to topology $D_2$ are shown in parentheses. The band structure of the points marked in blue are show in Fig. \ref{dist_bandas}.
}
\label{phase_Diagram}
\end{figure}

 \begin{figure}
\includegraphics[width=9cm]{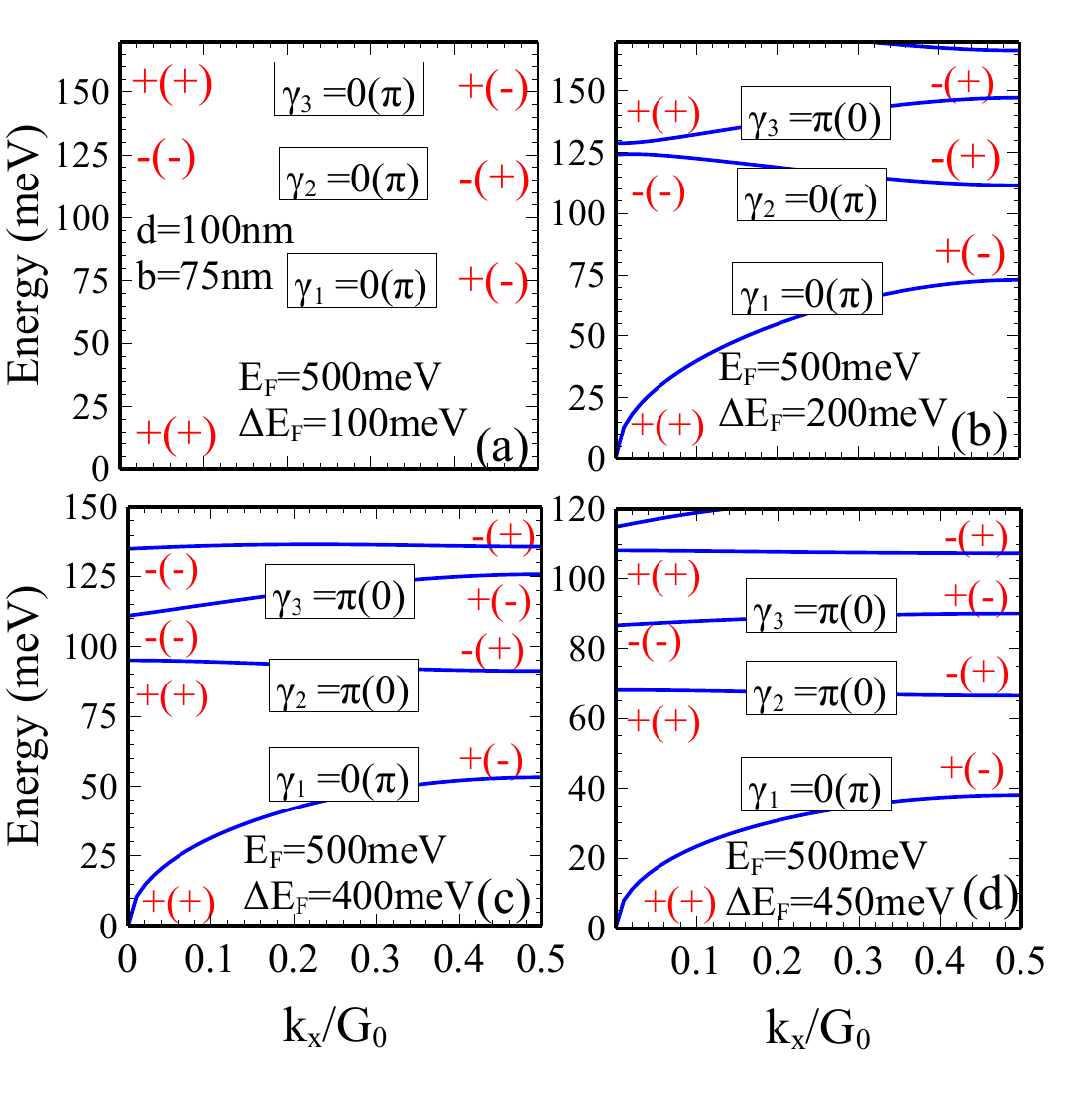}
\caption{
Plasmon band structures for a Kronig-Penney like model (see Fig. \ref{Plasmon_band}), with $d$=100nm, $b$=75nm and different values of $E_{F}^{(1)}$ and $E_{F}^{(2)}$, corresponding to the blue points marked in Fig. \ref{phase_Diagram}.
The parities of the wavefunctions for different plasmon subbands
at the points $\Gamma$ and $M$  are indicated in red.
We also show the Zak phase of the different subbands as obtained from Eq. \ref{gamma}. The values without/with parenthesis correspond to topologies $D_1$ and $D_2$, respectively.
}
\label{dist_bandas}
\end{figure}

Fig. \ref{phase_Diagram} illustrates the Zak phases of the three lowest energy subbands in the $D_1$ and $D_2$ topologies for different values of the
Kronig Penney lattice parameters.
To understand this behavior, in Fig. \ref{dist_bandas} we plot the band structure and the parity at the $\Gamma$ and $M$  wavevectors, for the points marked in blue in Fig.\ref{phase_Diagram}.
Since the $D_1$ and $D_2$ topologies correspond to locating the center of coordinates at different points within the unit cell, the subband Zak phases have opposite values in these states, as required by Eq. \ref{xntogamma}. The Zak phase of the lowest energy subband in the topology $D_1$ ($D_2$) takes the value $\gamma_1 = 0$ ($\pi$) across the entire range of parameters studied. In contrast, higher energy subbands exhibit transitions between different values of the Zak phases as the parameters of the Kronig-Penney lattices vary.

In the range of parameters we study, the lowest energy mode  disperse from the $\Gamma$ point ($k=0$) along the superlattice axis direction
as $\sqrt{k}$,
indicating that the main contribution to the  plasmon wavefunction, Eq.\ref{realwf} corresponds to ${\bf G}$=0, and the
plasmons propagate as a plane wave, $e^{i k x}$,
along the superlattice axis, and the parity of the state at the $\Gamma$ point is always 1.
Thus the Zak phase of the lowest energy subband is determined by the parity of the state at $X$ point, which in turn is fixed by the sign of the potential mixing the unperturbed plasmon states at momenta  $- G_0/2$ and $G_0 /2$ (see Eq.\ref{Eff_H}.)
In the topologies  $D_1$ and $D_2$, the origin of coordinates is located at the maximum and minimum  of the Fermi energy, respectively.
Therefore, for $D_1$
the sign of the Fourier component $E_{F,1}$ is positive, while for $D_2$ it is negative.
It follows that
the parity of  the lowest subband state at the $M$ point is positive in the $D_1$  topology, and is  negative for $D_2$.
As a consequence,  the Zak phase of the lowest energy subband,    $\gamma _1$,  is zero for $D_1$ (unit cell centered at maximum of the Fermi energy), and is $\pi$ for $D_2$ (unit cell centered at minimum of Fermi energy.)

For weak Fermi energy modulation, the parities of the states at the $\Gamma$ and $M$ points for the second subband are given by the signs of $E_{F,1}$ and $-E_{F,2}$, respectively.  From Eq. \ref{gamma}, this implies
$e ^{i \gamma _2}$=$ - {\rm sgn} ( E_{F,1}  E_{F,2} )$.  This suggests a Zak  phase transition at $b/d$=1/2, independent of $\Delta E_F$. However, as illustrated in Fig. \ref{phase_Diagram}, this only occurs for $\Delta E_F \to 0$. The discrepancy arises because the coupling to higher energy states results in a crossing of subbands from high energies at the $\Gamma$ point.
(Note, for example, the exchange of parities for states at the $\Gamma$ point for the second and third subbands, $\Gamma_2$ and $\Gamma_3$, in Fig. \ref{dist_bandas}(b)-(c).)  This endows the Zak phase boundary in Fig. \ref{phase_Diagram}(b) with a non-trivial  $b/d$ dependence.
The topology of the third band, $\gamma_3$, depends on the parities of the wavefunctions at  $\Gamma _3 $ and $M _3$.  Similar to the case of the second subband, under weak modulation, these parities are given by  the signs of the Fourier components of the Fermi energy that create  the gaps, so that one expects $e ^{i \gamma _3}$=$ - {\rm sgn} ( E_{F,2}  E_{F,3} )$ for small $\Delta E$. This induces Zak phase transitions at values of $b/d$=1/3, 1/2 and 2/3. Again, as the value of $\Delta E_F$ increases, the coupling with higher energy plasmon states becomes important, moving the Zak phase transitions to larger values of $b/d$.

\section{Edge and Interface Gap States.}
\subsection{Edge States.}
\begin{figure}
\includegraphics[width=8cm]{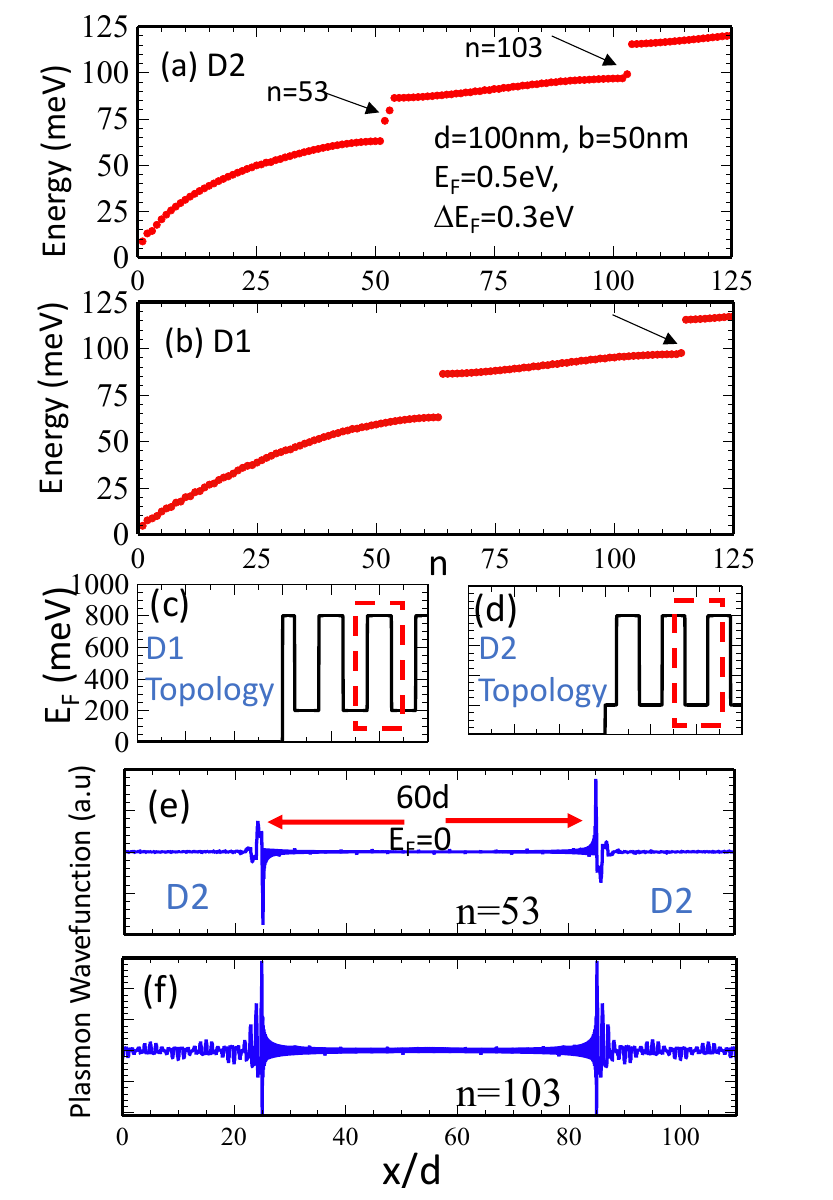}
\caption{(a) Energy spectrum of a superlattice with a unit cell consisting of 50 unit cells $d=100$nm of a KP-like plasmonic crystal in the  $D_2$
topology, separated by a region with zero Fermi energy that is 6000 nm wide. The other parameters of the system are indicated in the figure.
(b) Same as (a) but for the plasmon  superlattice in the state $D_1$.
(c) and (d): Fermi energy profile at the interface between the region with zero Fermi energy and the KP plasmonic crystal in the $D_1$ and $D_2$ topologies, respectively. The unit cells for $D_1$ and $D_2$ are highlighted  by a  red rectangle.
(e) Plasmon wave function corresponding to the state $n$=53 that appears inside the first energy gap in (a).
(f): Plasmon wave function for the state $n$=103, which appears in the second gap in (a).
}
\label{Edge}
\end{figure}

A fundamental property of systems with non-trivial topology is the appearance of in-gap states localized at a boundary of the system.  In the present context, the
topological state associated with a given gap is characterized by a  $\textit{Z}_2$ topological index,  Eq. \ref{Chern_index}. Such topological indices are very stable against small modifications of the bands (provided the underlying Hamiltonian still retains inversion symmetry). Changes to the topological index occur through topological phase transitions, which occur when gaps in the band structure close.  When the system has a boundary, there can effectively be an abrupt
change as one moves from the sample interior into the vacuum, for which there are no accessible states, making it effectively topologically trivial.
The change of a topological invariant at an edge induces the in-gap states \cite{Jackiw:1976aa,Su:1979aa}.

In the case of plasmonic crystals the vacuum corresponds to a region with zero Fermi energy, which does not support plasmons.
An interface between a plasmonic crystal and empty space may be simulated using a superlattice in which each unit cell includes a slab containing many unit cells of the plasmonic crystal, alongside a wide region with zero Fermi energy.
Both the slab and the vacuum region should be sufficiently wide to minimize coupling between states at the edges.

Fig. \ref{Edge}(a) and Fig.\ref{Edge}(b) illustrate the energy levels of a superlattice, with unit cell containing 50 unit cells of the plasmonic crystal with $d=$100nm in the topologies $D_2$ and $D_1$ respectively, with an adjacent region of zero Fermi energy 5000nm wide.  Note that in these systems, the topology is dictated by the form of the $E_F$ profile at the edge, which drops to zero at an inversion symmetry center of the Hamiltonian for the corresponding infinite system.
Figures \ref{Edge}(c) and (d) show the Fermi energy profile at the interface of a KP-like plasmonic crystal in topologies $D_1$ and $D_2$ (respectively) and the vacuum.

In Fig.\ref{Edge}(a), in the lowest energy gap, two states emerge, each corresponding to one of the two boundaries within the unit cell.
These two states are not fully degenerate because the finite size of the superlattice unit cell allows non-negligible interaction between the edge states.
These states arise  because the $\textit{Z}_2$ topological invariant in the first energy gap in the $D_2$ phase is $1$, in contrast with the vacuum space that is topologically trivial (topological invariant 0).
The wavefunctions of the in-gap states are localized at the edge of the sample. This is illustrated in Fig. \ref{Edge}(e), which shows the plasmon wave function corresponding to one of the states in the gap.
In the second energy gap one also finds a state. This arises due to the difference in the $\textit{Z}_2$ invariant between the second \textbf{gap} of the $D_2$ topology,
which remains at 1, and the vanishing topological invariant of the vacuum.
Only one state is visible in this gap due to the significant interaction between states localized at different edges. This causes a large enough level repulsion to shift one of the states into the continuum.
The effect is evident in Fig. \ref{Edge}(f), where the plasmon wave function of the state in the second energy gap is illustrated, and has significant weight in the plasmonic crystal region.

In the topology $D_1$ of the plasmonic crystal, the lowest energy subband hosts trivial topology, and no states appear in the lowest energy gap at the boundary of the crystal, as shown in Fig. \ref{Edge}(b). However, in the $D_1$
topology, the Zak phase of the second subband is $\pi$, leading to the emergence of a state in the second gap.  Note for the parameters used in the calculation, this resides barely outside the continuum.

\subsection{Interface States.}
States in the gap not only emerge at the boundary of a finite-size insulator with non-trivial topology but also at the interface between two insulators with different topological invariants $\nu$.  In this realization, such an interface arises when system is cut at two different inversion points, the finite segment removed, and the remaining semi-infinite systems are stitched together.

\begin{figure}
\includegraphics[width=8cm]{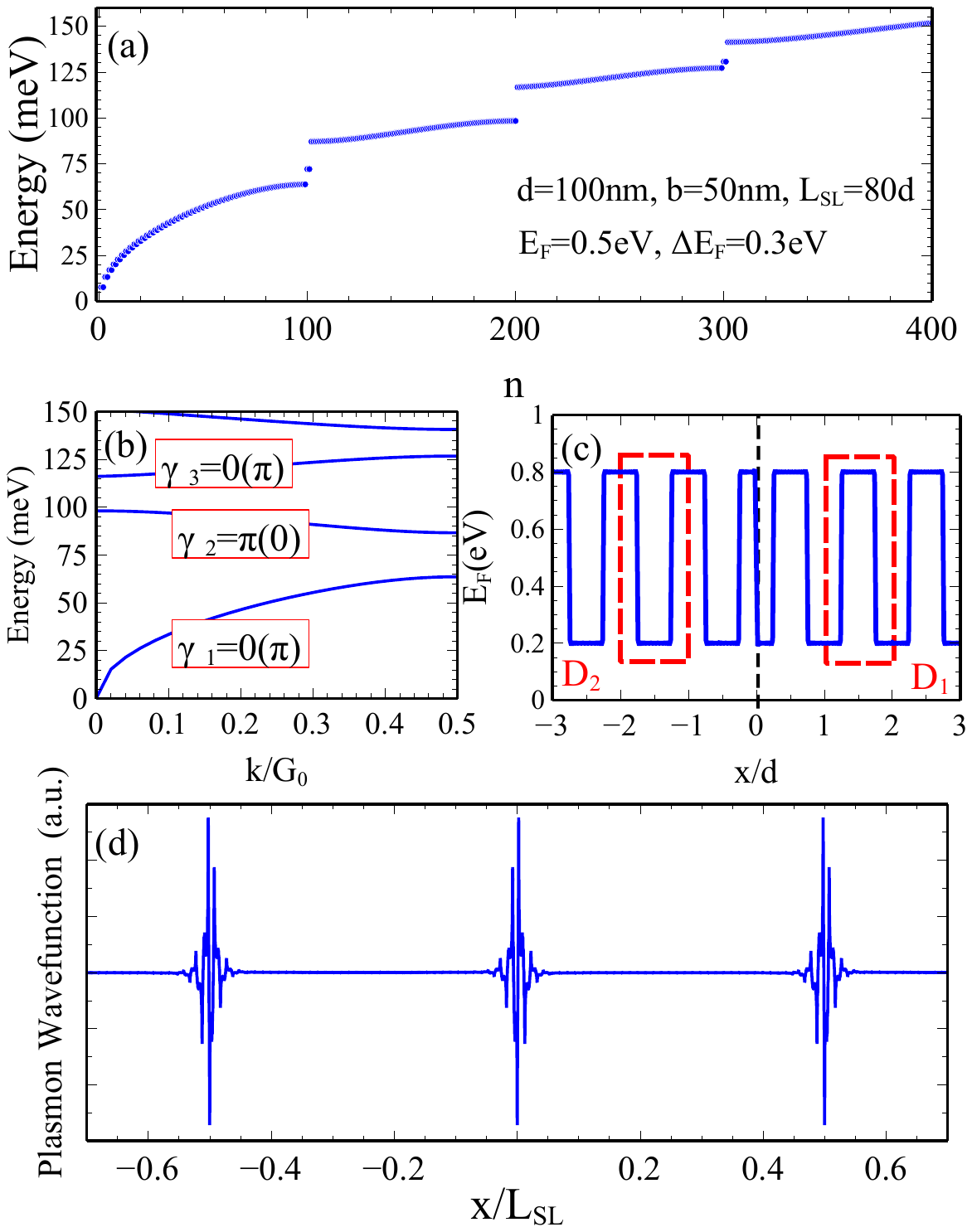}
\caption{
(a) The energy spectrum as a function of eigenstate number for a superlattice with a unit cell composed of 40 periods of a KP-like plasmonic crystal in the $D_1$
topology and another 40 periods in the $D_2$ topology.  The unit cell of the supercell is  $L_{SL}=80d$.
In-gap states appear in the first and third gaps but not
in the second gap.
(b) Band structure for the plasmonic crystal and subband Zak phases in the $D_1$
state. The Zak phases for the $D_2$ state are shown in parentheses.
(c) Interface between the $D_1$ and $D_2$ topologies, with the unit cells of both  highlighted in red.
(d) Plasmon wavefunction for an interface state within the first gap.
The parameters of the KP-like plasmonic crystal are displayed in the inset of (a).
}
\label{SL}
\end{figure}

To demonstrate this, we performed calculations on a superlattice with a large unit cell composed of 40 unit cells of a KP plasmonic crystal with $d=$100nm in the $D_1$ topology, and another with 40 unit cells in the $D_2$ topology.
Fig.\ref{SL}(c) shows the resulting interface, as well as the unit cells of both topologies.
In the spectrum of the large superlattice, plasmon states appear in the gaps between the first and second subbands, and between the third and fourth subbands, while no in-gap states appear between the second and third subbands.

This can be understood by examining the band structure of the KP-like plasmonic crystal shown in Fig. \ref{SL}(b).
When the energy gap is between the first and second subbands, the $\textit{Z}_2$ topological indices are $\nu_1=0$ for $D_1$ and $\nu_1=1$ for $D_2$. This change in the $\textit{Z}_2$ topological invariant leads to the appearance of plasmon states within the gap.
Similarly, for the gap between the third and fourth subbands, the $\textit{Z}_2$ topological invariants are also opposite: $\nu_3=1$ for $D_1$ and $\nu_3=0$ for $D_2$. This difference in topological indices leads to the appearance of in-gap plasmon states as well.
Conversely, for the gap between the second and third subbands, both $D_1$ and $D_2$ share the same $\textit{Z}_2$ topological invariant, $\nu_2=1$. In this case, there is no change in topology at the interface, and no edge states appear.

\section{Su-Schrieffer-Heeger Plasmonic Crystal.}
The Su-Schrieffer-Heeger (SSH) Hamiltonian
was introduced to describe  the transport properties of polyacetylene trough the motion of solitonic defects \cite{ Su:1979aa,Heeger:1988aa}, and
serves as a paradigm \cite{Hasan:2010aa,Cayssol:2021aa,Velasco:2017aa,McCann:2023aa,Perez-Gonzalez:2019aa}  for topology in condensed matter systems. Experiments across  different systems have confirmed  some of the topological properties predicted for the SSH model \cite{Meier:2018aa,Kiczynski:2022aa,Atala:2013aa,Cooper:2019aa}.

\begin{figure}
\includegraphics[width=8cm]{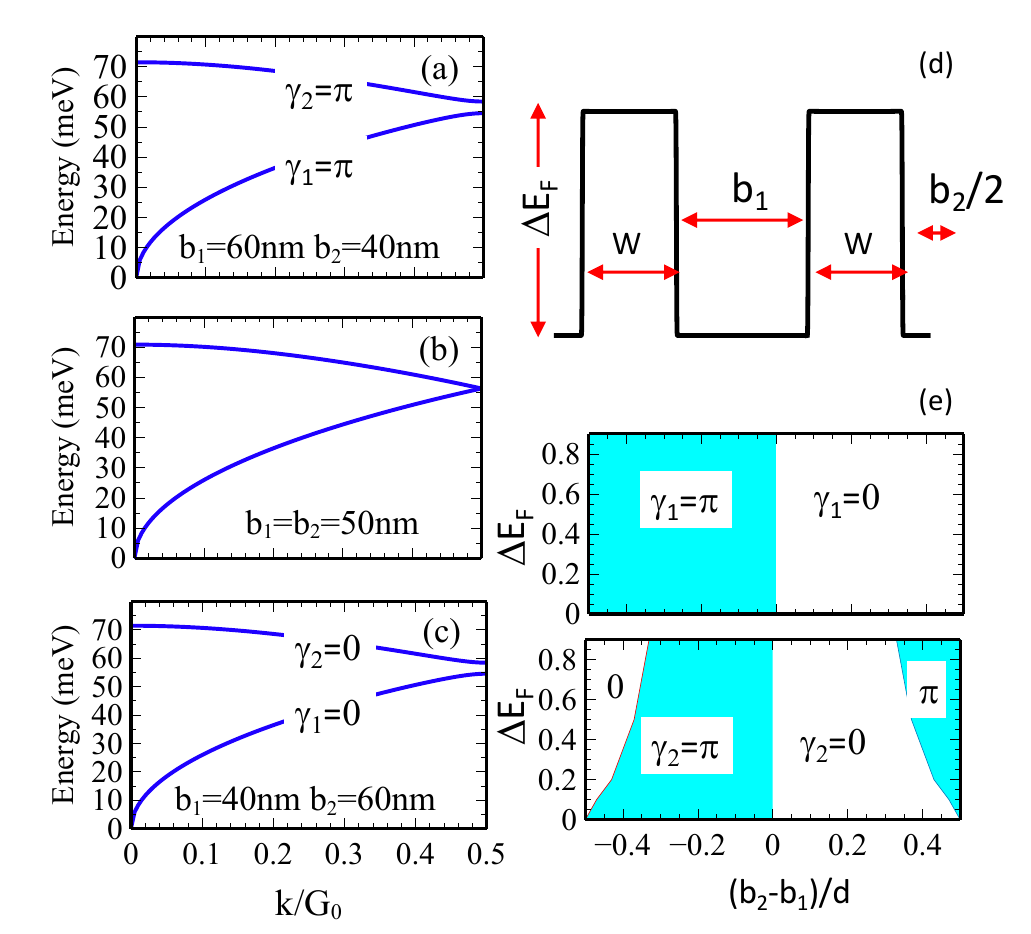}
\caption{ (a)-(c) Plasmon band structures and subband Zak phases for a SSH-like plasmonic crystal, with  unit cell shown schematically  in (d).
In (b), $b_1$=$b_2$ and there is no gap between the subbands.
(a) and (c) display  the bands for positive and negative values of $b_1-b_2$, respectively. The band energies are identical
but their topologies are opposite.
(e) Zak phases of the two lowest energy plasmon bands as functions of  $b_1-b_2$ and $\Delta E _F$.
Parameters used in these calculations are $d=200nm$, $W=500nm$, $b_1+b_2$=100nm, $E_F$=0.5eV. For the band structures (a)-(c), $\Delta E_F$=0.25meV.
}
\label{SSH}
\end{figure}

The SSH model is a tight-binding Hamiltonian containing  two identical atoms per unit cell, with two alternating hopping parameters, $t_1$ and $t_2$. The system has spatial inversion symmetry with symmetry points located at the centers of the two bonds.
The electronic spectrum consists of two bands separated by a gap proportional to the difference $t_1$-$t_2$.
The Wannier centers of the bands are  located in  the middle of the bond with the largest hopping parameter, $t_1$. The system exhibits two non-identical topologies, depending on how the inversion symmetry is realized in the Hamiltonian: either the parameter $t_1$ or $t_2$ may describe inter-site hopping within a unit cell, which is analogous to choosing which inversion center is used to realize inversion symmetry in the Hamiltonian.
For topology  $D_1$,  this intra-cell hopping parameter is chosen to be the larger of the two ($t_1$), and the bands have Zak phases equal to zero. Topology
$D_2$,  the intra-cell hopping is $t_2$, and  the Wannier center is effectively shifted to the unit cell edge, which the $t_1$ hopping effectively crosses. In this case the
Zak phases of the bands are $\pi$.  

For plasmons, an SSH model can be simulated with a unit cell containing two equal regions of width $W$
and Fermi energy $E_F$+$\Delta E_F$,
separated  by  alternating regions of  lower Fermi energy, $E_F$-$\Delta E_F$ and  thickness
$b_1$  and $b_2$ (see Fig.\ref{SSH}(d).) In an analogy with a single particle problem, the regions with the higher Fermi energy act as ``wells'' for the plasmons, while the regions with lower Fermi energy serve as ``barriers''
The coupling between the high Fermi energy regions decreases as the barrier thickness increases, and vice versa. In this way, the difference
$b_1$-$b_2$ is analogous to the difference $t_1$-$t_2$ in the SSH tight-binding model.

Fig. \ref{SSH}(e) shows the Zak phases for the SSH-like plasmonic crystal as a function of $b_2$-$b_1$ and $\Delta E_F$.
In the calculation, the origin  of the unit cell is located at the midpoint of the  region with low Fermi energy and width $b_1$. Except for very large values of $|b_2-b_1|$ and $\Delta E_F$, the Zak phase diagram is similar to that of the SSH tight-binding model. When $b_2 > b_1$ the Wannier center is
located at the center of the unit cell, $\bar x$=0, and the Zak phases of the lowest energy subbands are zero. By contrast, for negative values of
$b_2$-$b_1$ the Wannier center is located in the region with width $b_2$, and the Zak phases are $\pi$ for both subbands.
A topological phase transition occurs at the line defined by $b_2 =b_1$.
Figures \ref{SSH} (a)-(c) present the  plasmon bands energies  and Zak phases  for  $b_2-b_1$ values less than zero, equal to zero and greater than zero, respectively.
A topological phase transition occurs as  $b_2-b_1$ changes sign, with  the energy bands crossing  and the associated gap  closing  when $b_2=b_1$, as expected.

The topology of the SSH-like plasmonic crystals can be understood in terms of the first two Fourier components of the Fermi energy,
\begin{eqnarray}
E_{F,1} & = &  -\frac {\Delta E_F}{\pi} \left (   \sin {\frac {\pi} d b_1} -   \sin {\frac {\pi} d b_2}   \right ),   \nonumber \\
E_{F,2} & = &  -\frac {\Delta E_F}{\pi} \left (   \sin {\frac {2 \pi} d b_1} +   \sin {\frac {2 \pi} d b_2}   \right ).
\end{eqnarray}
As discussed in Section \ref{Zak1D}, for moderate values of $\Delta E_F$ the Zak phases of the lowest energy subbands are
$e ^{i\gamma _1}$=${\rm sgn} (E_{F,1}) $ and
$e ^{i \gamma _2}$=$ - {\rm sgn} ( E_{F,1}  E_{F,2} )$.
Combining these expressions with the forms of $E_{F,1}$ and $E_{F,2}$, one finds precisely
the topologies for bands of the SSH-like plasmonic crystal described above,
with oth low-energy subbands having the same Zak phase.


\section{Summary}
Starting from the quantization of plasmons in Drude-like doped graphene, we have constructed a Hamiltonian that describes the coupling among quantum plasmons with different momenta, driven by a periodic modulation of carrier density in the graphene sheet. The Hamiltonian contains terms that describe the scattering of plasmons due to the modulation, as well as non-conserving terms that involve the creation or annihilation of plasmon pairs.
We have diagonalized this Hamiltonian for the case of a one-dimensional modulation, and, because we are interested in unidirectional plasmonic waveguides, only focused on plasmons with vanishing momentum transverse to the modulations, $k_y$=0.
The modulation creates energy bands and gaps that are proportional to the Fourier components of the Fermi energy modulation.

When the modulation possesses spatial inversion symmetry, and the Hamiltonian is written in an inversion-symmetric way, the topology of each plasmon band is characterized by a Zak phase, 0 or $\pi$.  This is connected to the center of mass of the band Wannier functions, which lie at center or edge of the real space unit cell, respectively.
In a Kronig-Penney-like modulation of the Fermi energy, we find that the Wannier function of the lowest energy plasmon band is centered in the region with higher carrier density. Therefore, when the origin of coordinates in the Hamiltonian is located at this point,
and the first Fourier component of the Fermi energy modulation, $E_{F,1}$ is positive,
the Zak phase of the lowest energy band is 0, indicating it is topologically trivial state. However, if the origin is placed at  the center of the lower density region,
and $E_{F,1} $ is negative,
the band acquires a non-trivial topology, with  the Zak phase equal to $\pi$.
Thus, in systems with inversion symmetry,  the topological state of the lowest energy sub-band is determined by the sign of the first Fourier component of the Fermi energy modulation.
By contrast, for higher energy plasmon sub-bands, the gaps and band topologies depend on higher order Fourier components of the Fermi energy. By varying  the modulation of the Fermi energy or the ratio between the linear sizes of the high and low density regions, one may drive multiple transitions between different topological states.

A topological $\textit{Z}_2$ invariant can be defined at energies within the gap between two subbands. This invariant is either zero or one, depending on whether the sum of the Zak phases of the lower energy subbands is  an even or odd  multiple of  $\pi$.
An important property of band topology is that  in-gap states should appear  at the interface between two topological states with different topological Z$_2$ invariants.
We verified this property by computing the plasmon states at an interface between two systems with the same Kronig-Penney modulation of the Fermi energy, but with unit cells centered either in the higher or lower Fermi energy regions, respectively.
We also found in-gap states at the interface between a system with Kronig-Penney modulation of the Fermi energy and a non-metallic region.

Finally, we studied a Kronig-Penney superlattice with a unit cell that simulates an SSH system. The unit cell consists of two equal high-density regions, alternating with low-density regions of different lengths. We find that the topological states of this plasmonic crystal are highly analogous to those of the SSH model.

{\it Acknowledgements} -- LB was supported by Grant PID2021-125343NB-I00 (MCIN/AEI/FEDER, EU). 
HAF acknowledges the support of the NSF through
Grant No. DMR-1914451, and thanks the Aspen Center
for Physics (NSF Grant No. PHY-2210452) for its hospitality.

\appendix

\section{Semiclassical calculation of plasmons in a 2D electron gas with periodic modulated conductivity.}
In this Appendix,
we derive the plasmon spectrum for a two-dimensional system with optical conductivity  that is periodic in real space,
\begin{equation}
\bm \sigma  (\rn,\omega )
= \sum _{\Gn}
\sigma (\Gn,\omega )
e^{i \Gn \rn} \, \, .
\end{equation}
Suppose an external potential acts on the  layer,
\begin{equation}
\phi_{ext} (\rn ,\omega)= e ^{i \qn \cdot \rn} \sum _{\Gn}  \phi _{ext}   (\qn+\Gn,\omega) e ^{i \Gn \cdot \rn}.
\end{equation}
The system screens this potential, creating density oscillations in the layer, $  \rho (\qn+\Gn,\omega)$.
This in turn  induces an electrical potential of the form
\begin{equation}
\phi _{ind} (\qn+ \Gn,\omega) = v(|\qn+\Gn|)
\rho (\qn+ \Gn,\omega) \, \, ,  \label{coul}
\end{equation}
where $v(q)=2\pi e/\epsilon_0\epsilon_d q$.
The current in the system  responds  to the total electric field,
\begin{equation}
\Jn (\rn, \omega  ) = \sigma  (\rn,\omega)  \, \left[  \En _{ext}   (\rn ,\omega)  + \En _{ind}   (\rn ,\omega) \right] ,
\end{equation}
where we have assumed a purely local response to the electric field, and an isotropic, scaler conductivity.
Writing this equation in reciprocal space yields
\begin{eqnarray}
 \Jn (\qn+\Gn ,  \omega ) =  -i \sum _{\Gn'} (\qn +\Gn') \, \sigma  (\Gn-
\Gn',\omega )   \nonumber \\  \times \left (
 \phi _{ext}  (\qn+\Gn',\omega) +  \phi _{ind}  (\qn+\Gn',\omega) \right) \,.
 \label{ohm}
\end{eqnarray}
Combining the continuity equation in reciprocal space,
\begin{equation}
(\qn+\Gn)  \cdot  \Jn (\qn+\Gn, \omega ) = \omega  \rho (\qn+\Gn,\omega) \,,
\label{cont}
\end{equation}
with Eqs. \ref{coul} and \ref{ohm} yields
\begin{eqnarray*}
 \phi  _{ind} (\qn+ \Gn, \omega  )   = -   \frac {i} \omega  v(|\qn+\Gn|)
\sum_{\Gn '} (\qn+\Gn) \cdot (\qn + \Gn ')  \nonumber \\
\!\!\!\!\!\!\!\!\!\!\!\! \times
\sigma (\Gn \! - \! \Gn ')
\left[
 \phi _{ind}(\qn+ \Gn',\omega)+  \phi _{ext}(\qn+ \Gn',\omega)
 \right] \, .
\end{eqnarray*}
This implies
\begin{eqnarray}
  \phi  _{tot} (\qn+ \Gn,\omega)  & = &  \phi  _{ext} (\qn+ \Gn,\omega)-  \frac {i} {\omega}  v(|\qn+\Gn|)
\nonumber \\
& \times &   \sum_{\Gn '}(\qn+\Gn) \! \cdot \! (\qn + \Gn ')
\sigma (\Gn  -  \Gn ')
\phi _{tot}(\qn+ \Gn',\omega) , \nonumber
\end{eqnarray}
where $\phi ^T_{tot}(\qn+ \Gn,\omega)   =    \phi _{ind}(\qn+ \Gn,\omega)+  \phi_{ext}(\qn+ \Gn,\omega)$ is the total
electric potential in the system. This equation defines a dielectric matrix (with reciprocal lattice vectors acting as the matrix indices) for the system,
\begin{eqnarray*}
\epsilon  (\qn&\!+\!& \Gn  ,   \qn+ \Gn ')
=
 \delta _{\qn+\Gn,\qn+\Gn'} \nonumber \\
& +&     \frac {i} \omega  v(|\qn+\Gn|)
(\qn+\Gn)\cdot(\qn + \Gn ')
\sigma  (\Gn \! - \! \Gn' )
\end{eqnarray*}
Zero eigenvalues of the dielectric matrix determine the plasmon frequencies as functions of $q$.

In the case of Drude-like metals, where the optical conductivity has the form $\sigma (\rn,\omega)=i \frac {D(\rn)} {\omega} $,
the dielectric constant takes the form
\begin{equation*}
\epsilon (\qn+\Gn,\qn+\Gn')= \ \delta _{\qn+\Gn,\qn+ \Gn'} -\frac 1 {\omega ^2} M(\qn+\Gn,\qn+\Gn '),
\end{equation*}
with
\begin{equation*}
M(\qn+\Gn,\qn+\Gn')=v(|\qn+\Gn|) (\qn +\Gn)\cdot(\qn+\Gn ') D(\Gn -\Gn '),
\end{equation*}
where $D(\Gn )$ is the Fourier transform of $D(\rn)$.

For a given momentum $\qn$  the dielectric matrix is diagonalized by the same matrix $U$ that diagonalizes $M$,
\begin{equation}
\sum_{\Gn'} \epsilon ( \qn +\Gn, \qn+ \Gn ') U_{\Gn',i} = U_{\Gn,i} \Lambda _i
\end{equation}
where the eigenvalues $\{ \Lambda _i \}$ are related with the eigenvalues of $M$, $\{ \lambda_i \}$, through
\begin{equation}
\Lambda _i =1- \frac {\lambda _i } {  \omega ^2} \, \, .
\end{equation}
Thus the frequencies of the \textbf{plasmon modes} are determined by the square roots of the eigenvalues of $M$.

The microscopic dielectric constant can be related to the observed macroscopic dielectric function
$\bar {\epsilon}  (\qn,\omega)$ by
\begin{equation*}
\bar {\epsilon} (\qn,\omega)= \frac 1 {  \left [  \left (\epsilon (\Gn, \Gn ')   \right ) ^{-1} \right ]_{\Gn=\Gn'=0}}.
\end{equation*}
A representative result for this quantity is shown in Fig. \ref{Plasmon_band}.
Experimentally observable plasmons \cite{Pisarra:2016aa} correspond to the zeros of $\bar {\epsilon} (\qn,\omega)$.
Numerically these appear at the maxima of
\begin{equation*}
{\rm Im} \left [  \left (\epsilon (\Gn, \Gn ')   \right ) ^{-1} \right ]_{\Gn=\Gn'=0}=\sum _i  U_{\Gn=0,i}
\frac {\eta \omega \lambda_i}{(\omega^2-\lambda_i  )^2+\eta ^2 \omega ^2}
U^{{-1}} _{\Gn=0,i},
\end{equation*}
where $\eta$ is a quasiparticle lifetime broadening.  Note that this last quantity can slightly shift the frequencies of the plasmons.

$\quad$

$\quad$

\section{Zak Phase in Systems with Spatial Inversion Symmetry}

The Zak phase is given by the expression

\begin{equation}
\gamma _n = i \int _{-\frac {\pi} d} ^{\frac {\pi} d} dk \langle \, { U}  _{n,k} (x) \,  \sigma _z  \, \partial _k { U}_{n,k} (x) \, \rangle
\label{B1}
\end{equation}
where ${ U}_{n,k} (x)$ is the cell-periodic part of the Bloch wavefunction,
\begin{equation}
{\Psi}_{n,k} (x)= e ^{i kx } { U} _{n,k} (x).
\end{equation}
\hfil\break
The Pauli matrix appears in the definition of the Zak phase because of the symplectic character of the Bogoliubov-Hopfield transformation
\cite{Shindou:2013aa,Goren:2018aa}.
Because the system is periodic, the inner product $\langle \, { U}  _{n,k} (x) \,  \sigma _z  \, \partial _k { U}_{n,k} (x) \,\rangle$ involves an integration over a single unit cell.
Then
\begin{widetext}
\begin{equation}
\gamma _n= i \int _{-\frac {\pi} d} ^{\frac {\pi} d} dk \langle \, { \Psi}  ^* _{n,k} (x) \,  \sigma _z  \, \partial _k { \Psi} _{n,k} (x) \, \rangle +
\int _{-\frac {\pi} d} ^{\frac {\pi} d} dk \langle \,{ \Psi} _{n,k} (x) \,  \sigma _z  \, x {\Psi} _{n,k} (x) \,\rangle.
\end{equation}
The origin of coordinates for the system is chosen at a center of inversion symmetry, and
the last term is zero because if this.
On the other hand systems with spatial inversion symmetry satisfy
\begin{equation}
{ \Psi} _{n,k} (x) = e ^{-i \phi_n(k) } \, \hat {\cal  I}  \,  { \Psi} _{n,-k} (x)\, ,
\end{equation}
where $\hat {\cal I} $ is the spatial inversion operator, and $\phi_n(k)$ is an arbitrary phase. Then
\begin{eqnarray}
\gamma_n & = &
i \int _0 ^{\frac {\pi} d} dk < { \Psi}  _{n,k} (x) \,  \sigma _z \, \partial _k { \Psi} _{n,k} (x) >
+
i \int _{-\frac {\pi} d}  ^0 dk < {\bm \Psi}  _{n,-k} (x) \,  \hat {\rm I} ^{\dagger} e ^{-i \phi(k) } \, \sigma _z \, \partial _k e ^{-i \phi_n(k) } \hat {\rm I} | {\bm \Psi} _{n,-k} (x) > \nonumber \\
& =&
i \int _0 ^{\frac {\pi} d} dk < {\Psi}  _{n,k} (x) \,  \sigma _z , \partial _k {\Psi} _{n,k} (x) >+
i \int _{-\frac {\pi} d} ^0 dk < { \Psi}  _{n,-k} (x) \,  \sigma _z \,\partial _k { \Psi} _{n,-k} (x) >
+ \int _{-\frac {\pi} d} ^0 dk \partial _k \phi _n(k) \nonumber \\
& = & i \int _0 ^{\frac {\pi} d} dk < { \Psi }  _{n,k} (x) \,  \sigma _z \,  \partial _k { \Psi}  _{n,k} (x) >-
i \int _{\frac {\pi} d} ^0 dk < { \Psi}   _{n,k} (x) \,  \sigma _z \, \partial _{-k} { \Psi}  _{n,k} (x) >
+ \int _{-\frac {\pi} d} ^0 dk \partial _k \phi _n(k) \nonumber \\
& = & i \int _0 ^{\frac {\pi} d} dk < {\Psi}  _{n,k} (x) \,  \sigma _z  \, \partial _k { \Psi} _{n,k} (x) >+
i \int _ 0 ^{\frac {\pi} d}  dk < {\Psi}  _{n,k} (x) | \sigma _z \, \partial _{-k} { \Psi} _{n,k} (x) >
+ \int _{-\frac {\pi} d} ^0 dk \partial _k \phi _n(k) \nonumber \\
& = & i \int _0 ^{\frac {\pi} d} dk < {\bm \Psi}  _{n,k} (x) | \sigma _z \partial _k {\bm \Psi} _{n,k} (x) >-
i \int _ 0 ^{\frac {\pi} d}  dk < {\bm \Psi}  _{n,k} (x) |\sigma _z \partial _{k} {\bm  \Psi} _{n,k} (x) >
+ \int _{-\frac {\pi} d} ^0 dk \, \partial _k \phi _n(k) \nonumber \\
&=& \phi_n(0)-\phi(-\frac {\pi} d) = \phi_n (0)-\phi_n(-\frac {G_n} 2)
\end{eqnarray}
In the second line above we have used the normalization condition
$\sum_{G_n}
\left[ ({\alpha ^n   _{k+G_n} } )^2 -(\beta ^{n }_{k+G_n})^2 \right] $=1.

In general the phase $\phi _n  (k)$ can take any value for an arbitrary value of $k$.  However, for a time reversal invariant momentum, $k_{T}$, which satisfies
$k_{T}=-k_{T}+ G_n$ , the phase should obey
\begin{equation}
{\Psi}  _{n,k_T} = e ^{i \phi_n (k_T)} \hat {\cal I} \, { \Psi}  _{n,k_T}= e ^{i \phi_n  (k_T)} \xi _n (k_T) { \Psi} _{n,k_T}
\end{equation}
where $\xi_n (k)$ is the parity eignevalue of the wavefunction at this wavevector.

Therefore, at time reversal invariant momentum points, the phase must take the \textbf{values 0 or  $\pi$}.  Specifically,
if $\xi _n (k_T)=1$, the phase is $\phi_n(k_T)=0$, and if $\xi _n (k_T)=-1$ the phase is $\phi_n(k_T)=\pi$. Thus,

\begin{equation}
\gamma _n = \phi_n(0)-\phi(-\frac {\pi} d) = \left\{
\begin{array} {l}0  \, \, {\rm  \, when \, states \, at}\,  k=0 \,  {\rm and} \, {k=G_0/2} {\rm \, have \, the \, same \, symmetry} \\
   \pi  {\rm \,\,  when \,  states \, at} \, k=0 \,  {\rm and}  {k=G_0/2} {\rm \, have \, opposite \,  symmetry}
 \end{array}\right\}.
 \end{equation}
This leads to the final result
 \begin{equation}
e^{i \gamma _n} = \xi _n (k=0) \, \xi _n (k= \frac {G_0}  2)
 \end{equation}
 which is presented as Eq. \ref{gamma} in the main text.
\end{widetext}

%

\vspace{0.5truecm}
\end{document}